\documentstyle[prl,aps,epsf]{revtex}
\begin{document}
\draft
\title{
Renormalization of the Coulomb interaction in one-dimensional electron systems}
\author{S. Bellucci $^1$ and J. Gonz\'alez $^2$  \\}
\address{
	$^1$Laboratori Nazionali di Frascati.
        INFN.
        P. O. Box 13. I-00044 Frascati. Italy.\\
        $^2$Instituto de Estructura de la Materia.
        Consejo Superior de Investigaciones Cient{\'\i}ficas.
        Serrano 123, 28006 Madrid. Spain.}
\date{\today}
\maketitle
\begin{abstract}
Renormalization group methods are used to study the low-energy
behavior of the unscreened Coulomb interaction in a
one-dimensional electron system. By applying a GW approximation,
a strong wavefunction renormalization is found in the model, as
well as renormalization of the Fermi velocity. This may have
significant effects depending on the filling level of the
system. The effective couplings remain bounded 
at arbitrarily low energies so that, in spite of
the long-range character of the interaction, the system still
falls into the Luttinger liquid universality class.

\end{abstract}

The recent experimental availability of single fullerene
nanotubes has renewed the interest in the study of
one-dimensional electron systems\cite{nano}.  
In one spatial dimension the
Luttinger liquid concept replaces the Fermi liquid picture, and
provides the paradigm of a system with strong electronic
corrrelations. While the fullerene tubules are genuine
one-dimensional systems and, therefore, appropriate to test the
consequences of the electronic correlations, it is also true for
the same reason that the bare interaction is not screened in the
case of isolated tubules. 
The analysis of the Luttinger liquid behavior is
usually made under the assumption of a local or short-range
interaction\cite{sol,hal,vt,gz}, 
but the effects that the unscreened Coulomb
interaction may have on the properties of the one-dimensional
system have not been completely assessed.

The bosonization approach, in particular, 
is one of the most elegant and
rigorous ways of describing the Luttinger liquid behavior.
However, the Coulomb interaction is singular at small momentum
transfer, and the usual
formulas that give the renormalized parameters like the Fermi
velocity $v_F$ in terms of the interactions $g_2 (q)$ and $g_4
(q)$ cannot be applied at arbitrarily small energy scales,
unless the Coulomb potential is screened by external
charges\cite{fisher}. 
The more physical reason for the
unfeasibility of the bosonization approach to the present
problem regards the construction of the ground state of the
system in terms of the density operators of the noninteracting
theory. Such state is normalizable provided that at large
momenta $\sqrt{q} g_2(q) /(2\pi v_F + g_4(q) ) \rightarrow 0$
\cite{hal},  while it
is unclear that this condition may be satisfied in the case of
the Coulomb interaction.

In this Letter we are going to use renormalization group (RG)
methods to find the low-energy effective theory of the $1/|x|$ 
interaction. 
The main
point that we want to address is the stability of the Coulomb
interaction in the RG framework, and whether it falls into
the Luttinger liquid universality class. In order to deal
conveniently with this question one has to introduce
the dynamical screening due to plasmons, what we
accomplish by implementing a GW approximation. The suitability
of this approximation in the study 
of one-dimensional systems has been recently shown in Ref.
\onlinecite{gw}.
The same approach has been also tested in the study of the crossover 
from Fermi liquid to Luttinger liquid behavior\cite{dicastro},
as well as in the study of singular interactions in dimension $1
< d \leq 2$ \cite{wen}.

To be more precise, we want to pose the problem of a
one-dimensional one-band model with
an
interaction hamiltonian
\begin{equation}
H_{int} = \frac{e^2}{8\pi } \int dx dx'\; \Psi^{+} (x) \Psi (x)
\frac{1}{|x - x'|} \Psi^{+} (x') \Psi (x')
\label{ham}
\end{equation}
where $\Psi (x)$ is the electron annihilation field operator.
The RG method is a sensible approach to deal with this problem
since the $1/|x|$ interaction potential (as well as the $\delta
(x)$ potential) gives rise to a marginal four-fermion interaction.
The scaling dimension of the electron field $\Psi
(x)$ is $- 1/2$ , in length units. This means that the interaction
hamiltonian in (\ref{ham}) scales appropriately, with a 
dimensionless coupling constant $e^2$ (in units in which $\hbar
= c = 1$), as the energy scale is 
reduced down to the Fermi level\cite{sh}.

However, the drawback in dealing with (\ref{ham}) is
that it contains a highly nonlocal operator, which makes unclear
the applicability of RG methods, usually devised to deal with a
set of local operators.
This problem can be circumvented by introducing a local
auxiliary field to propagate the Coulomb interaction. 
The interaction potential in real space $1/|x - x'|$ corresponds
to a singular interaction $\sim \log|q|$ in momentum space.
Therefore, it makes sense to focus on the electron
states that lie close to any of the two Fermi points of the band
at $\pm k_F$ , and to consider two respective left $(L)$ and
right $(R)$ branches with
linear dispersion relation for such relevant states. In
practice, one enforces this approximation by considering the
excitations that have less energy than a given 
bandwidth cutoff $E_c$ .
The hamiltonian can be written in the form
\begin{equation}
H  =  i v_F \int dx \; \left( \Psi^{+}_R (x) \partial_x \Psi_R (x)
 -  \Psi^{+}_L (x) \partial_x     \Psi_L (x)  \right)
 +   e \int dx \; \left( \Psi^{+}_L (x) \Psi_L (x) +
\Psi^{+}_R (x) \Psi_R (x) \right) \; \phi (x)
\label{ham2}
\end{equation}
where the $\phi (x)$ field propagates the interaction
\begin{equation}
i \langle T  \phi (x,t) \; \phi (x',t') \rangle = \frac{1}{4\pi}
\delta (t - t') \frac{1}{|x - x'|}
\end{equation}

In writing (\ref{ham2}) we have neglected backscattering
processes that connect the two branches of the dispersion
relation. This is justified, in a first approximation,
as for the Coulomb interaction the processes with
small momentum transfer have much more strength than those with
momentum transfer $ \sim 2k_F $ . The backscattering processes
give rise, however, to a marginal interaction, that should be
taken into account after renormalization of the dominant forward
scattering channel.

We may think of $\phi (x)$ as the scalar potential in 
three-dimensional quantum
electrodynamics.  However, the differences with that theory in
the present case are notorious since the propagation of $\phi
(x)$ is that of a genuine field in three spatial dimensions,
while the electrons are confined to one dimension. In general,
one may expect a better infrared behavior in the present model.
The propagator of the $\phi (x)$ field in momentum space can be
read from the relativistic expression\cite{landau}, 
after sending the speed
of light to infinity,
\begin{equation}
i \langle T  \phi (x,t) \; \phi (x',t') \rangle = \int \frac{dq
d\omega}{(2 \pi)^2} \int \frac{dq_y dq_z}{(2 \pi)^2}
\frac{\mbox{\Large $e^{i q (x - x')}$ }
\mbox{\Large $e^{-i \omega (t - t')}$ }
}{ q^2 + q_y^2 + q_{z}^{2} - i \epsilon }
\label{prop}
\end{equation}
The usual one-dimensional propagator $\sim
\log(|q|/\Lambda ) $ is recovered from (\ref{prop}) 
upon integration of the
dummy variables $q_y$ and $q_z$. We remark that the ultraviolet
cutoff $\Lambda $ for excitations along the $y$ and $z$
transverse directions is needed when projecting the
three-dimensional interaction down to the one-dimensional
system. 

The usefulness of the representation (\ref{prop}) can be
appreciated in the renormalization of the model at the one-loop
level. We study the scaling behavior of the irreducible
functions as the bandwidth cutoff $E_c$ is sent towards the
Fermi level, $E_c \rightarrow 0$. The self-energy to the
one-loop order is
\begin{equation}
i \Sigma (k,0) = i e^2 \int^{E_c}_{-E_c} \frac{ dp}{2\pi }
 \int^{+\infty}_{-\infty} \frac{ d\omega_p}{2\pi }
\frac{v_F (p + k)}{\omega_p^2 + v_F^2 (p + k)^2} \int \frac{dp_y
dp_z}{(2\pi)^2} \frac{1}{p^2 + p_y^2 + p_z^2}
\label{pert}
\end{equation}
The limit $k \rightarrow 0$ has to be taken carefully in this
expression, by first combining the two denominators with the use
of Feynman parameters\cite{landau}. Finally we get
\begin{eqnarray}
i \Sigma (k,0) & = &  \frac{i}{4\pi} \frac{e^2}{v_F}
\int_{0}^{1} du \frac{1}{\sqrt{u}} \int_{-E_c}^{E_c}
\frac{dp}{2\pi} \int_{-\infty}^{\infty} \frac{d\omega_p}{2\pi}
\frac{v_F k} {\omega_p^2 + p^2 + v_F^2 k^2 u(1-u) }   \nonumber
\\ &  \approx  & i \frac{e^2}{4\pi^2} k \; \log E_c
\label{ren}
\end{eqnarray}
The term linear in $k$ in $\Sigma (k,0)$ represents a
renormalization of the Fermi velocity, which grows upon
integration of the high-energy modes. Obviously, there is no
correction linear in $\omega_k$ renormalizing the electron
wavefunction at the one-loop level. This is consistent with the
fact that the integration of high-energy modes at $\sim E_c$
does not renormalize the three-point vertex $\Gamma $ . 
We want to stress the difference of the logarithmic
renormalization of $v_F$ with respect to the usual {\em finite}
corrections due to a short-range interaction. The nontrivial
scaling of $v_F$ is a genuine effect of the long-range
Coulomb interaction, which also takes place in higher
dimensions\cite{np}.

Incidentally, the above computation exemplifies how
the Ward identity that ensures the integrability of the
Luttinger model does not hold in the present case. The Ward
identity is a relation between the electron Green function $G(p,
\omega_p)$ and the three-point vertex $\Gamma (p, \omega_p ; k,
\omega_k)$ at a given branch\cite{sol}. For the right-handed
modes, for instance, it is
\begin{equation}
\Gamma (p, \omega_p ; k, \omega_k) =
 \frac{G^{-1} (p, \omega_p) - G^{-1} (p-k, \omega_p - \omega_k)}
{\omega_k - v_F k}
\label{wi}
\end{equation}
By focusing on the singular dependences on the bandwidth
cutoff $E_c$, one can check that (\ref{wi}) is already violated in our
model to first order in perturbation theory. Actually, the
dependence of the vertex $\Gamma $ on the variables $(k,
\omega_k)$ of the external interaction line is 
\begin{equation}
i \Gamma (p, \omega_p ; k, \omega_k) \approx -i
\frac{e^2}{4\pi^2} \frac{k \; \log \Lambda }{\omega_k - v_F k}
\end{equation}
We notice that the Ward identity would be satisfied if the
scaling could be implemented simultaneously in the transverse
ultraviolet cutoff $\Lambda $ and the bandwidth cutoff $E_c$.
However, even in a real system the scaling in $\Lambda $ gets
locked by the finite cross section of the wire, while only the
scaling in the longitudinal direction operated by $E_c$ is
allowed. In this respect, the gauge invariance of
quantum electrodynamics is broken by the
anisotropy of the electron system, as felt by the propagation of
the three-dimensional electromagnetic field.

The renormalization of $v_F$ at the one-loop level is not, in general, 
a sensible effect from the physical point of view, since the
propagator of the $\phi (x)$ field is drastically modified by
the quantum corrections. Its one-loop self-energy is 
given by the sum of two respective particle-hole
diagrams with modes in the left and the right branch of the
dispersion relation. The result for both fermion loops is finite,
and they combine to give
\begin{equation}
i \Pi (k, \omega_k) = i \frac{e^2}{\pi} \frac{v_F k^2} {v_F^2
k^2 - \omega_k^2 }
\label{pol}
\end{equation}
Taking into account these particle-hole processes leads to a
modified propagator of the $\phi (x)$ field
\begin{equation}
i \langle \phi (k,\omega ) \; \phi (-k,-\omega ) \rangle =
 1 / \left( -\frac{2 \pi}{\log(|k|/\Lambda) } + \Pi (k,\omega ) \right)
\label{prop2}
\end{equation} 
The expression (\ref{prop2}) provides a sensible
approximation for the scalar propagator, as it takes into account
the plasmons in the model. Quite notably, the result
(\ref{prop2}) turns out to be the exact propagator in
the Luttinger model with short-range interactions\cite{dl}.

Our approach is that of using the scalar propagator
(\ref{prop2}) in the renormalization of the Fermi velocity and
the electron wavefunction. For this purpose we compute the
electron self-energy by replacing the interaction in Fig. \ref{one} by
the dressed interaction (\ref{prop2})
\begin{equation}
i \Sigma (k, i \omega_k) = i \frac{e^2}{2\pi } \int^{E_c}_{-E_c} 
\frac{dp}{2\pi }
 \int^{+\infty}_{-\infty} 
\frac{d\omega_p}{2\pi } \frac{1}{i (\omega_p + \omega_k) - v_F (p +
k) } \frac{\log(|p|/\Lambda) }{1 - \frac{e^2}{2\pi^2 } \frac{v_F p^2}{v_F^2
p^2 + \omega_p^2 } \log(|p|/\Lambda) }
\label{selfe}
\end{equation}
This corresponds to taking the leading order in a $1/N$
expansion, in a model with $N$ different electron flavors.  In
our case, such approximation to the self-energy is also
justified since it takes into account, at each level in
perturbation theory, the most singular contribution at small
momentum transfer of the interaction. Due to the
cancellation of fermion loops with more than two interaction vertices which
still takes place in the same way as in the Luttinger model,
the representation
(\ref{selfe}) for the self-energy only misses
the effects of vertex and self-energy corrrections, but these
are treated consistently in the RG framework.

The only contributions in (\ref{selfe}) depending on the
bandwidth cutoff are terms linear in $\omega_k$ and $k$. In
this respect, 
it is worth mentioning that, although the usual perturbative approach
gives rise to poles of the form $k^2/(\omega_k - v_F k)$ in the
self-energy\cite{dl}, these do not arise in the GW approximation. 
There is no infrared catastrophe at $\omega_k \approx v_F k$,
because of the correction in the slope of the plasmon dispersion
relation with respect to its bare value $v_F$. 
The contribution at small $k$ has to be dealt with particular 
care, anyhow, as the interaction
is not invariant under shifts in the momentum transfer. The
result that we get for the renormalized electron propagator is
\begin{eqnarray}
G^{-1}(k,\omega_k) & = &  Z^{-1}_{\Psi} \; (\omega_k - v_F
  k)  - \Sigma (k,\omega_k)  \nonumber   \\ & \approx &
Z^{-1}_{\Psi} \; (\omega_k - v_F  k) -
Z^{-1}_{\Psi} \; (\omega_k - v_F k)  \int^{E_c} 
 \frac{dp}{|p|} \left( 1 - \frac{1+f(p)}{2 \sqrt{f(p)}}
\right)                  \nonumber               \\ &  &  -
Z^{-1}_{\Psi} \; k \; \frac{e^2}{4\pi^2 }\int^{E_c} 
 \frac{dp}{|p|} 
   \frac{\sqrt{f(p)} - 4/3  +  1/ \left( 3  f(p)^{3/2} \right)  }
  { \left( 1-f(p) \right)^2  }
\label{gren}
\end{eqnarray}
where $f(p) \equiv 1 - e^2 \log(|p|/\Lambda) /(2 \pi^2 v_F) \;  $ and
$Z^{1/2}_{\Psi}$ represents the scale of the
bare electron field compared to that of the cutoff-independent
electron field
\begin{equation}
\Psi_{bare}(E_c) = Z^{1/2}_{\Psi} \Psi
\end{equation}

In the RG approach, we
require the cutoff-independence of the renormalized Green
function, since this object leads to observable quantities in
the quantum theory. For this purpose,
the quantities $Z_{\Psi}$ and $v_F $ have to be promoted to
cutoff-dependent
effective parameters, that reflect the behavior of the quantum
theory as $E_c \rightarrow 0$ and more states are integrated out
from high-energy shells of the band. We get then the RG flow equations
\begin{eqnarray}
E_c \frac{d}{dE_c }\:  \log \: Z_{\Psi}(E_c)  & = & 
   \frac{1+f(E_c)}{2 \sqrt{f(E_c)}} - 1      \label{zflow}   \\
E_c \frac{d}{dE_c } \: v_F (E_c)  & = &  - \frac{e^2}{4\pi^2}
    \frac{\sqrt{f(E_c)}  -  4/3  +  1/ \left(3 f(E_c)^{3/2} \right)}
    {\left( 1-f(E_c) \right)^2} 
\label{vflow}
\end{eqnarray}

One can check that Eq.
(\ref{vflow}) leads to an enhancement of the Fermi velocity at
low energies, which goes in the direction of screening the
effective interaction. Eq. (\ref{zflow}) also corresponds to a
sensible effect, as the leading behavior is that of suppressing
the electron quasiparticle weight. It can be shown that, within
the same level of the $1/N$ approximation, the three-point
vertex only gets the cutoff dependence given by the wavefunction
renormalization in (\ref{zflow}). This means that the electron
charge is not renormalized at low energies in our local field
theory framework. The behavior of the effective interaction is,
therefore, completely encoded in Eq. (\ref{vflow}).

The RG equation for the effective coupling constant $g \equiv
e^2 /(4\pi^2 v_F)$ is
\begin{equation}
E_c \frac{d}{dE_c } g(E_c) = \frac{1}{4(\log \: E_c)^2} \left(
 \sqrt{f(E_c)} - \frac{4}{3} + \frac{1}{3f(E_c)^{3/2}} \right)
\label{gflow}
\end{equation}
This equation actually controls all the physical properties of
the electron system and, in particular, the vanishing of the electron
quasiparticle weight. In our model the effective
coupling constant $g$ displays marginal behavior. However, the
logarithmic corrections to scaling 
in (\ref{gflow}) reduce the flow in the
infrared. That equation could still admit a solution of the form
$g \sim - g_0 / \log \: E_c $, but this is not realized 
in the present model as the
equation $2 g_0 = \sqrt{1+g_0} - 4/3 + 1/(3\sqrt{(1+g_0)^3})$
does not have any real solution. The flow of $g (E_c)$ 
given by (\ref{gflow})
seems to be arrested close to some fixed-point value, after
which it becomes insensitive to further scaling in the infrared.
A plot of the flow for different values of the 
bare coupling constant is given in Fig. \ref{two} .

We see therefore that the Coulomb interaction remains long-ranged
in the low-energy effective theory, while the effective coupling
has a stable flow in the infrared. This means that the
one-dimensional system falls into the Luttinger liquid
universality class. It can be seen from (\ref{zflow}), in
particular, that the electron quasiparticle weight vanishes in
the low-energy effective theory. The fast wavefunction
renormalization shown in Fig. \ref{three} seems to imply a non-algebraic
behavior of the electron propagator with regard to the frequency
and momentum dependence, what is reminiscent of similar features
found for some correlation functions in the bosonization
approach\cite{schulz2}.

Moving to the possible implications for real systems, the degree
of renormalization of $v_F$ may depend, in general, on the details of the
band structure, as the result (\ref{vflow}) has been obtained
by using the straight form of the polarizability
(\ref{pol}) for all values of the momentum. The issue of the 
renormalization of $v_F$ is important since it
implies a reduction in the strength
of additional interactions in the system, whether short or
long-range, as the effective couplings are all given by the
couplings in the interaction hamiltonian divided by $v_F$.

In particular, one may
envisage real systems where the Fermi level is close to the top or
the bottom of the band, so that the polarizability significantly
deviates from (\ref{pol}). For very low filling, for instance,
it vanishes at large momentum transfer like $k_F /E_c$, meaning
that the RG flow should be computed according to the
renormalization given by (\ref{ren}), until the dynamical
screening due to plasmons takes place. We have represented in
Fig. \ref{two} the flow of the effective coupling under these
circumstances. In this situation there may be a significant
renormalization of $v_F$ that affects every effective interaction
in the model. They flow towards a nonvanishing fixed-point,
in any event, since at sufficiently low energies there has to
be a point in which one recovers the Luttinger liquid
regime, where the RG flow equations (\ref{vflow}) and
(\ref{gflow}) apply. The significant renormalization of $v_F$
may also be present in small chains, contributing to explain
the insulator-metal transition by the effect of the Coulomb
interaction observed in the exact diagonalization of finite
rings\cite{poil}.

In conclusion, we have treated the long-range Coulomb
interaction in a one-dimensional electron system by applying RG
methods on a $1/N$ truncation of the theory. 
We have seen that the Coulomb interaction 
remains long-ranged in the low-energy effective theory, and that
this feature is compatible with Luttinger liquid behavior of the
system. This does not seem to
support any phase transition with regard to the strength of the
interaction. However, significant renormalizations of all the
effective interactions may arise depending on the band structure,
what could be a relevant matter in developping the phenomenology of real
one-dimensional electron systems and, in particular, the fullerene
nanotubes\cite{fisher}.

\begin{figure}
\caption{Diagram contributing to the electron self-energy. The
double line stands for the interaction propagator computed in
the RPA, according to Eq. (10) in the text.}
\label{one}
\end{figure}

\begin{figure}
\caption{Flow of the effective coupling constant for different
bare values, computed in GW approximation (full lines) and
without the dynamical screening of plasmons (dashed lines).}
\label{two}
\end{figure}

\begin{figure}
\caption{Wavefunction renormalization for bare
coupling constant $g = 5.0$ (thick line) and $g = 1.0$ (thin line).}
\label{three}
\end{figure}

\newpage

\par
\centering
\epsfbox{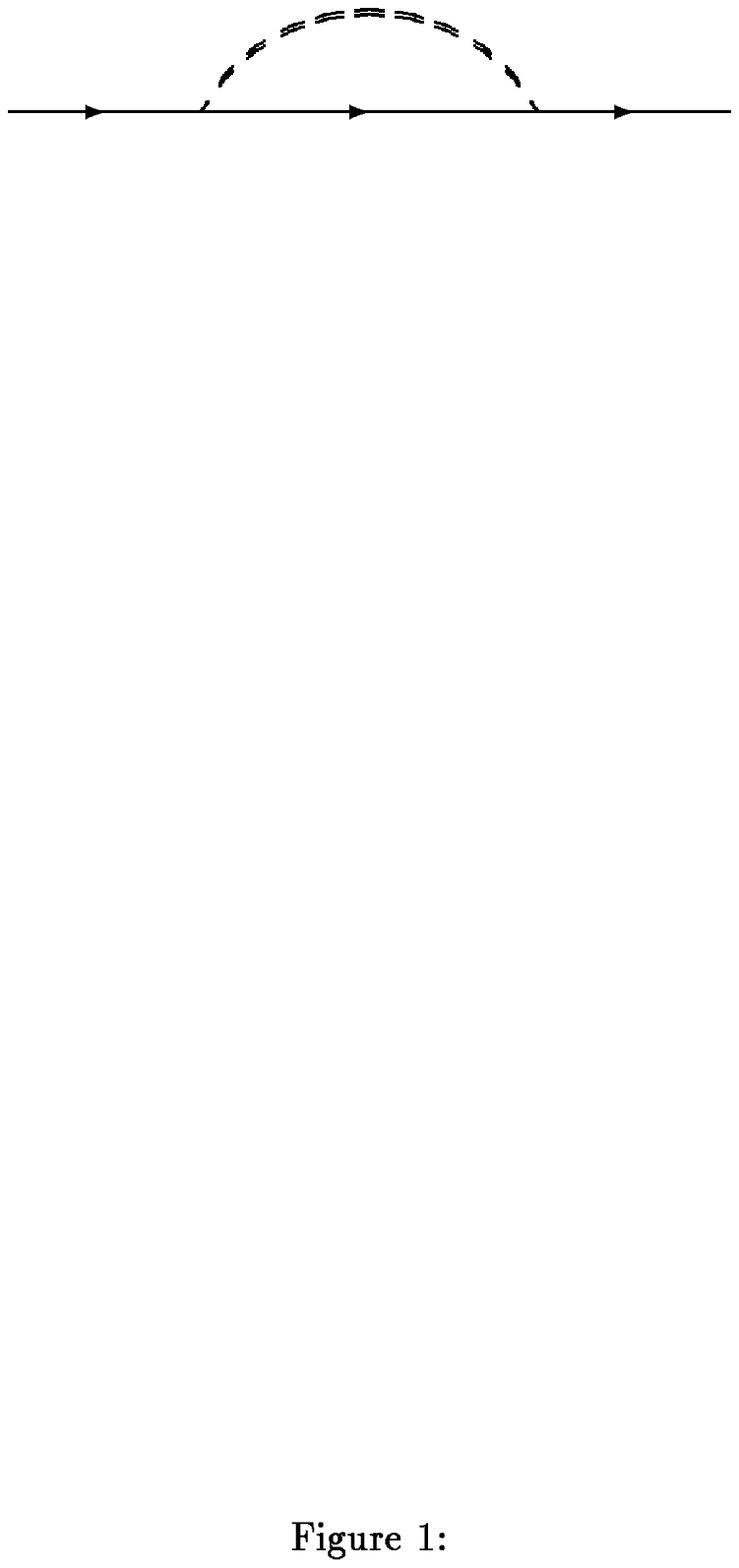}
\par

\newpage

\par
\centering
\epsfbox[0 800 332 1132]{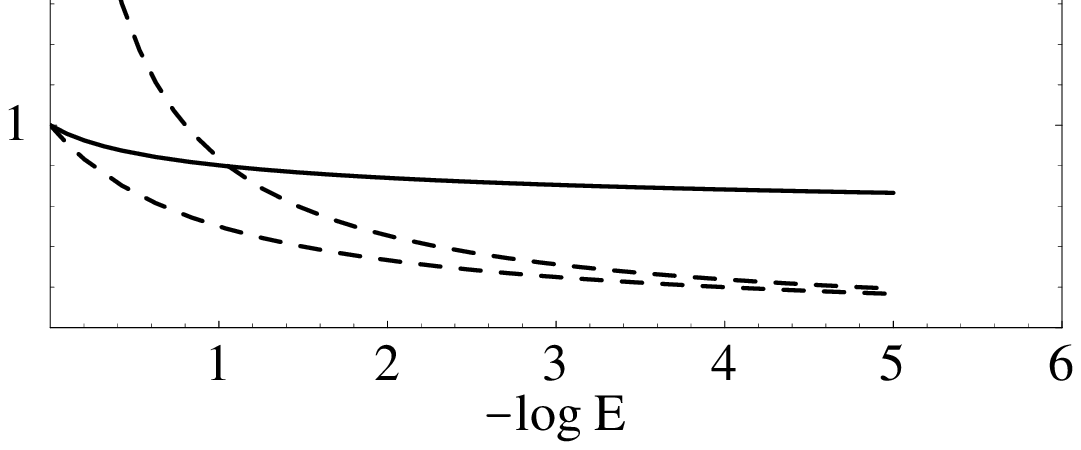}
\par

\newpage

\par
\centering
\epsfbox[ 0 800 324 1124]{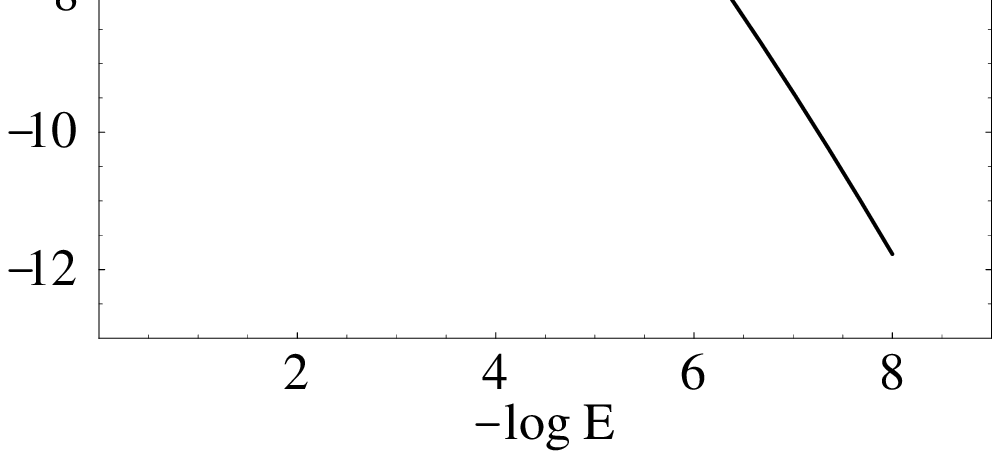}
\par

\end{document}